\documentclass[aps,prl,preprint,tightenlines,superscriptaddress,showpacs,byrevtex]{revtex4}
\usepackage{graphicx} 
\usepackage{dcolumn}  
\usepackage{amsmath}
\usepackage{tabularx}
\usepackage{hhline}
\usepackage{epstopdf}
\usepackage{rotating}
\usepackage{color}
\usepackage{charter}
\usepackage{bm}
\usepackage{subfigure}

\graphicspath{{ps}}

\begin{document}

\newcolumntype{Z}{>{\raggedleft\arraybackslash}X}
\newcolumntype{Y}{>{\center}X}
\newcolumntype{W}{>{\raggedright\arraybackslash}X}


\preprint{\vbox{ \hbox{   }
						\hbox{BELLE Preprint 2009-14}
						\hbox{KEK Preprint 2009-12}
}}

\title{ \quad\\[1.0cm]  Measurement Of  $|V_{ub}|$ From Inclusive Charmless Semileptonic $B$ Decays}


\begin{abstract}
We present the partial branching fraction for inclusive charmless semileptonic $B$ decays and the corresponding value of the CKM matrix element $|V_{ub}|$, using a multivariate analysis method to access $\sim$90\% of the $B \to X_u \ell \nu$ phase space. This approach dramatically reduces the theoretical uncertainties from the $b-$quark mass and non-perturbative QCD compared to all previous inclusive measurements.  The results are based on a sample of 657 million $B \bar B$ pairs collected with the Belle detector.   We find that $\Delta \mathcal{B}(B \to X_u \ell \nu; p^{*B}_{\ell}>1.0~{\rm GeV}/c)=1.963 \times (1 \pm  0.088_{\rm stat.}  \pm  0.081_{\rm sys.}   )\times 10^{-3}$. Corresponding values of $|V_{ub}|$ are extracted using several theoretical calculations.
\end{abstract}

\affiliation{Budker Institute of Nuclear Physics, Novosibirsk}
\affiliation{Chiba University, Chiba}
\affiliation{University of Cincinnati, Cincinnati, Ohio 45221}
\affiliation{T. Ko\'{s}ciuszko Cracow University of Technology, Krakow}
\affiliation{The Graduate University for Advanced Studies, Hayama}
\affiliation{Hanyang University, Seoul}
\affiliation{University of Hawaii, Honolulu, Hawaii 96822}
\affiliation{High Energy Accelerator Research Organization (KEK), Tsukuba}
\affiliation{Hiroshima Institute of Technology, Hiroshima}
\affiliation{Institute of High Energy Physics, Chinese Academy of Sciences, Beijing}
\affiliation{Institute of High Energy Physics, Vienna}
\affiliation{Institute of High Energy Physics, Protvino}
\affiliation{Institute for Theoretical and Experimental Physics, Moscow}
\affiliation{J. Stefan Institute, Ljubljana}
\affiliation{Kanagawa University, Yokohama}
\affiliation{Institut f\"ur Experimentelle Kernphysik, Universit\"at Karlsruhe, Karlsruhe}
\affiliation{Korea University, Seoul}
\affiliation{Kyungpook National University, Taegu}
\affiliation{\'Ecole Polytechnique F\'ed\'erale de Lausanne (EPFL), Lausanne}
\affiliation{Faculty of Mathematics and Physics, University of Ljubljana, Ljubljana}
\affiliation{University of Maribor, Maribor}
\affiliation{University of Melbourne, School of Physics, Victoria 3010}
\affiliation{Nagoya University, Nagoya}
\affiliation{Nara Women's University, Nara}
\affiliation{National Central University, Chung-li}
\affiliation{National United University, Miao Li}
\affiliation{Department of Physics, National Taiwan University, Taipei}
\affiliation{H. Niewodniczanski Institute of Nuclear Physics, Krakow}
\affiliation{Nippon Dental University, Niigata}
\affiliation{Niigata University, Niigata}
\affiliation{University of Nova Gorica, Nova Gorica}
\affiliation{Novosibirsk State University, Novosibirsk}
\affiliation{Osaka City University, Osaka}
\affiliation{Panjab University, Chandigarh}
\affiliation{RIKEN BNL Research Center, Upton, New York 11973}
\affiliation{Saga University, Saga}
\affiliation{University of Science and Technology of China, Hefei}
\affiliation{Seoul National University, Seoul}
\affiliation{Sungkyunkwan University, Suwon}
\affiliation{University of Sydney, Sydney, New South Wales}
\affiliation{Tata Institute of Fundamental Research, Mumbai}
\affiliation{Toho University, Funabashi}
\affiliation{Tohoku Gakuin University, Tagajo}
\affiliation{Tohoku University, Sendai}
\affiliation{Department of Physics, University of Tokyo, Tokyo}
\affiliation{Tokyo University of Agriculture and Technology, Tokyo}
\affiliation{IPNAS, Virginia Polytechnic Institute and State University, Blacksburg, Virginia 24061}
\affiliation{Yonsei University, Seoul}
  \author{P.~Urquijo}\affiliation{University of Melbourne, School of Physics, Victoria 3010} 
  \author{E.~Barberio}\affiliation{University of Melbourne, School of Physics, Victoria 3010} 
  \author{I.~Adachi}\affiliation{High Energy Accelerator Research Organization (KEK), Tsukuba} 
  \author{H.~Aihara}\affiliation{Department of Physics, University of Tokyo, Tokyo} 
  \author{K.~Arinstein}\affiliation{Budker Institute of Nuclear Physics, Novosibirsk}\affiliation{Novosibirsk State University, Novosibirsk} 
  \author{A.~M.~Bakich}\affiliation{University of Sydney, Sydney, New South Wales} 
  \author{K.~Belous}\affiliation{Institute of High Energy Physics, Protvino} 
  \author{V.~Bhardwaj}\affiliation{Panjab University, Chandigarh} 
  \author{M.~Bischofberger}\affiliation{Nara Women's University, Nara} 
  \author{A.~Bozek}\affiliation{H. Niewodniczanski Institute of Nuclear Physics, Krakow} 
% ...
  
  \collaboration{The Belle Collaboration}


\maketitle

\tighten

{\renewcommand{\thefootnote}{\fnsymbol{footnote}}}
\setcounter{footnote}{0}

The comparison of the Cabibbo-Kobayashi-Maskawa (CKM) matrix element $|V_{ub}|$~\cite{ckm}, which determines one of the sides of the Unitarity Triangle, and the CP angle $\phi_1$~\cite{notation} is one of the crucial tests of the Yukawa sector of the Standard Model.  The angle $\phi_1$ is directly sensitive to potential CP-violation beyond the Standard Model and can be measured with high precision without theoretical input.  In contrast, $|V_{ub}|$ is insensitive to new physics as it is determined by tree-level weak  decays but relies on input from  theoretical QCD calculations.  Currently measurements of $\phi_1$ and $|V_{ub}|$ are not entirely consistent \cite{CKMfit}.

Experimentally  $|V_{ub}|$  can be measured from semileptonic decays either using an exclusive hadronic final state, or by considering the inclusive rate, summing over all hadronic final states subject to some kinematical constraints. The two approaches involve different experimental and theoretical tools.  Currently there is a $\sim2\sigma$ discrepancy~\cite{HFAG} in the respective world averages, indicating an incomplete understanding of these tools.  The extraction of $|V_{ub}|$ is a challenge for both theory and experiment.  The primary difficulty in measuring $|V_{ub}|$ with high precision in inclusive $B \to X_u \ell \nu$ decays is suppressing the background from $B \to X_c \ell \nu$, which is $50$ times larger.   All measurements to date have applied kinematic selection criteria to achieve separation from $B \to X_c \ell \nu$ decays. These required the use of the theoretical parameterizations called shape functions (SF) to describe the unmeasured regions of phase space.  Here we report a measurement of the partial branching fraction of $B \to X_u \ell \nu$ decays with a lepton momentum threshold of 1 GeV/$c$ using a multivariate data mining technique, and derive  values of $|V_{ub}|$ using several theoretical calculations \cite{BLNP,DGE,Gambino:2007rp}.   This approach accesses $\sim$90\% of the $B \to X_u \ell \nu$ phase space,  the greatest reach of any inclusive $|V_{ub}|$ measurement \cite{Bizjak:2005hn, Aubert:2007rb, Aubert:2005im}.  This is a major milestone in the measurement of $|V_{ub}|$, as we can rely on the well tested theory used to describe  $B \to X_c \ell \nu$ transitions, minimizing the dependence on a SF.  Thus this is the single most precise determination of $|V_{ub}|$, and provides a valuable new direction for $|V_{ub}|$ determinations by addressing previously irreducible theoretical uncertainties.

The measurement is made by fully reconstructing one $B$ meson ($B_{\rm tag}$) decaying to a fully hadronic final state, and identifying the semileptonic decay of the other $B$ meson ($B_{\rm sig}$) by the presence of a high momentum electron or muon.  The data were collected with the Belle detector \cite{Belle} at the asymmetric-energy KEKB $e^+e^-$ collider~\cite{KEKB}. The results presented in this Letter are based on a sample of $657 \times 10^6 B\bar{B}$ pairs collected at the $\Upsilon(4S)$ resonance (on-resonance).  An additional $68$ fb$^{-1}$ data sample taken at $60$ MeV below the $\Upsilon(4S)$ resonance (off-resonance) is used to perform subtraction of background arising from the continuum $e^+ e^- \to q \bar q$ process ($q$ $=$ $u$, $d$, $s$, $c$).  

The $B_{\rm tag}$ candidates are reconstructed following the procedure of Ref.~\cite{matsumoto}, in hadronic modes that determine their charge, flavour, and momentum. For each selected candidate, we calculate the beam-energy constrained mass, $M_{\mathrm{bc}} = \sqrt{(E^*_{\mathrm{beam}})^2-(p^*_B)^2}$, and the energy difference, $\Delta E = E^*_B-E^*_{\mathrm{beam}}$, where $E^*_{\mathrm{beam}}$, $p^*_B$ and $E^*_B$ are the beam energy, the reconstructed $B$ momentum and the reconstructed $B$ energy in the $\Upsilon(4S)$ rest frame, respectively.   In events containing multiple $B$ meson candidates, the candidate with the smallest $\chi^2_B$, defined in Ref.~\cite{matsumoto}. The $B_{\rm tag}$  purity of this sample is $25$\% ($30$\%) for $B^+$ ($B^0$) tags~\cite{cc}.  Events with   $M_{\mathrm{bc}} \in (5.27, 5.29)$ $\mathrm{GeV}/c^2$ , $|\Delta E|< 0.05~\mathrm{GeV}$, and $\chi^2_B<10$ are considered for further analysis.  True $B\overline{B}$ events for which the reconstruction of $B_{\rm tag}$ is not correct are treated as background (referred to as combinatorial background). This background peaks in the signal region of $M_{\rm{bc}}$.   We derive the shape of the combinatorial background from Monte Carlo (MC) as in Ref.~\cite{Urquijo:2006wd}, with the yield normalized to the on$-$resonance data $M_{\mathrm{bc}}$ sideband ($M_{\mathrm{bc}} \in (5.20, 5.25)$ $\mathrm{GeV}/c^2$) after the subtraction of non-$B \overline{B}$ (continuum) backgrounds.  The continuum background is scaled by the integrated on- to off-resonance luminosity ratio, taking into account the cross-section difference.  There are $1167329\pm5412_{\rm stat.}$ $B$  candidates in the  signal region ($N_{\rm{tag}}$),  after continuum and combinatorial background subtraction.  

Electron and muon candidates decaying from $B_{\rm sig}$ are required to originate from near the interaction vertex and pass through the barrel region of the detector, corresponding to an angular acceptance of $\theta_{\rm lab} \in (35^{\circ},125^{\circ})$  ($\theta_{\rm lab} \in (25^{\circ},145^{\circ})$) for electrons (muons), where $\theta_{\rm lab}$ denotes the polar angle of the lepton candidate with respect to the direction opposite to the positron beam. We exclude tracks used in the reconstruction of the $B_{\rm{tag}}$ and multiple reconstructed tracks generated by   low-momentum particles spiraling in the drift chamber.  We consider the lepton with the highest momentum in the $B$~rest frame to be prompt.  The lepton identification efficiencies and the probabilities to misidentify a pion, kaon or proton as a lepton have been measured as a function of the laboratory momentum and angles.  The average electron (muon) identification efficiency and hadron misidentification rate are $97$\% ($90$\%) and $0.7$\% ($1.4$\%), respectively, over the full phase space.  In $B^+$~tagged events, we require the lepton charge to be consistent with a prompt semileptonic decay of $B_\mathrm{sig}$. In $B^0$~events, we make no requirement on the lepton charge. For semileptonic $B$ decays to electrons, we partially recover the efficiency loss due to bremsstrahlung as in Ref.~\cite{Urquijo:2006wd}. The lepton momenta are calculated in the $B$ meson rest frame ($p^{*B}_{\ell}$).  Events with leptons from $J/\psi$ decays, photon conversions, and $\pi^0$ decays are rejected using the invariant mass of prompt lepton candidates in combination with an oppositely charged lepton; for electron candidates  additional photons are included in the veto calculation.

The $B \to X_u \ell \nu$ selection criteria are based on a non-linear multivariate analysis technique,  the Boosted Decision Tree (BDT) method~\cite{Hocker:2007ht}, which takes into account various observables to form one event classification variable.  A total of $17$ discriminating variables are used to form a BDT classifier, separating $B \to X_u \ell \nu$ decays from other kinds of $B$ decays.  These include quantities based on: the kinematics of the candidate semileptonic decay; discrete quantities such as the number of kaons; and quantities correlated to the quality of the event reconstruction, such as $M_{\rm bc}$.  A description of the highest discriminating quantities follows. The absolute value of event net charge is found to be correlated to track multiplicity, which tends to be higher for  $b\to c$ transitions. The kinematic variables associated to the hadronic current, $M_X$ and $P_+$ (invariant mass, and energy-momentum of the hadronic system, $X_u$, respectively) are calculated from the measured momenta of all charged tracks and neutral clusters that are not associated to $B_{\rm tag}$ reconstruction or used as lepton candidates.   The lepton current four-momentum is calculated as $q = p_{\Upsilon(4S)}- p_{B_{\rm tag}} - p_X$.  Missing momentum attributed solely to prompt neutrinos should have a missing mass consistent with zero. Thus we calculate the missing mass squared, $m^2_{\rm miss}$, of the events from the missing four-momentum $P_{\rm miss}$.  The missing momentum is estimated from the four-momenta of the tag-side $B$ and all reconstructed charged particles and photons that pass selection criteria on the signal side: $P_{\rm miss} = P_{\Upsilon(4S)} - P_{B_{\rm tag}} - \sum_{\rm charged}P- \sum_{\rm neutral}P$.   To reduce contamination from $B \to D^* \ell \nu$ events, we search for low momentum pions from $D^{*+} \to D^0 \pi^+$ and calculate the momentum of the $D^{*+}$ and missing mass squared, $m_{{\rm miss}(D^*)}^2 \equiv (P_{B_{\rm sig}}-P_{D^*}-P_{\ell})^2$. The presence of kaons in semileptonic $B$ meson decay is usually an indication of a $b \to c$ transition, although $b \to u$ decays with kaons from $s \bar{s}$ popping  in the final state have been observed.  Such decays are far less abundant than the charm cascade production of kaons,  thus the number of charged kaons and $K_{\rm S}^{0}$ mesons are considered in the multivariate analysis. We set an event selection threshold criterion for the BDT-classifier that is optimized with respect to both the systematic uncertainty from the background normalization fit and phase space dependent theoretical uncertainties.  We set a lower threshold on $p^{*B}_{\ell}$ of $1.0$~GeV/$c$.

\begin{figure}[htb!]
\begin{center}
\includegraphics[width=.80\textwidth]{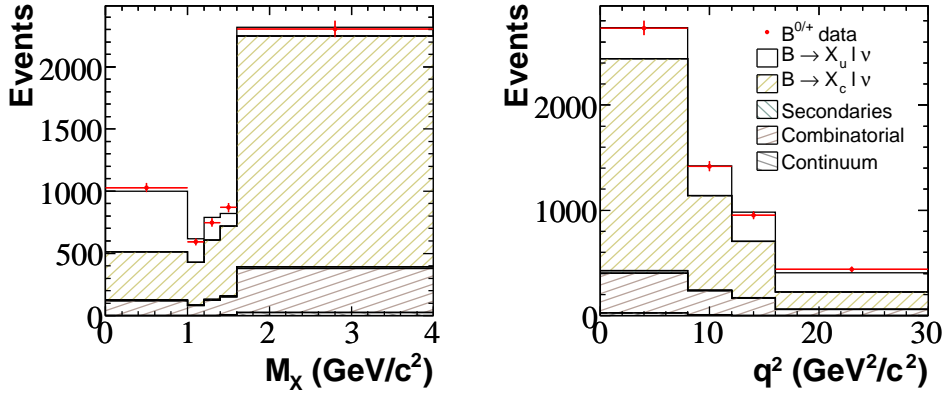}
\end{center}
\caption{Projections of the $M_X - q^2$ fit in bins of $M_X$ (left) and $q^2$ (right).}
\label{fig:mx_fit}
\end{figure}

The backgrounds that remain after the BDT selection criteria are subtracted as described below. The continuum and combinatorial backgrounds follow the $N_{B \bar B}$ determination procedure described earlier in this Letter. All remaining backgrounds arise when the fully reconstructed $B$ is correctly tagged, but the decay is either a charmed semileptonic $B$ decay, a secondary decay process that produced a high momentum lepton or is a misidentified hadron. The shapes of the charmed semileptonic $B$ decay contribution, described in detail in Ref.~\cite{Urquijo:2006wd}, and the secondary contribution, are determined from MC simulation.  We estimate the overall normalization of these remaining backgrounds by fitting the observed inclusive spectra to the sum of the MC simulated signal and background contributions, after continuum and combinatorial background subtraction.  There are three free parameters in the fit, corresponding to the yields of:  $B \to X_u \ell \nu$; $B \to X_c \ell \nu$; and secondaries and fakes. The fit is performed in two dimensional bins of $M_X$ versus $q^2$ for $4684 \pm 85$ input events,  with a lepton momentum requirement of $p^{*B}_{\ell}>1.0$~GeV$/c$.  The fit results in a good agreement between data and MC, with a $\chi^2$ of $24$ for $17$ degrees of freedom (Figure~\ref{fig:mx_fit}). A total of $1032 \pm 91$ events remain after background subtration.
We measure the partial branching fractions, combining the spectra from $B^+$ and $B^0$ semileptonic decays with the $1.0$~GeV/$c$ lepton momentum threshold.  The expression for the partial branching fraction is  $\Delta\mathcal{B}= (N_{b \to u}^{\Delta}/(2\epsilon_{b \to u}^{\Delta}N_{\rm tag}))(1-\delta_{\rm rad})$, where $N_{b \to u}^{\Delta}$ and $\epsilon_{b \to u}^{\Delta}$ are the signal yield and signal efficiency for the region, $\Delta$ ($p^{*B}_{\ell}\ge1.0$~GeV/$c$), $N_{\rm{tag}}$ is the number of tagged $B$ events and $\delta_{\rm rad}$ denotes QED corrections.  The overall efficiency is $22.2$\%, determined from the fully reconstructed signal MC,  reweighted at the generator level in bins of $p_{\ell}$, $P_+$, $M_X$ and $q^2$ following the prescription in this Letter.    The QED correction is $1.4$\% of the branching fraction, obtained using Ref.~\cite{PHOTOS}.  The various contributions to the systematic error on the partial branching fraction are described below.  

To estimate the particle identification and reconstruction uncertainties, events with electrons and muons are reweighted and  kaons, pions and photons are randomly removed according to their respective measured uncertainties.

The MC sample used for the signal $B \to X_u \ell \nu$ events is a hybrid mix of inclusive and exclusive contributions. Resonant semileptonic $B$ decays to $\pi$, and $\rho$ and $\omega$ modes are modeled with form factors calculated in Ref.~\cite{LCSR} and Ref.~\cite{BALL}, respectively, with branching fractions set to the world averages ~\cite{HFAG}. Decays to  $\eta$ and $\eta^\prime$ have form factors derived from Ref.~\cite{ISGW2} and branching fractions set to the world averages~\cite{HFAG}. The form factors and branching fractions of the unmeasured resonant components are predicted by Ref.~\cite{ISGW2}. The branching fractions of the resonant $B \to X_u \ell \nu$ final states have been varied by $\pm 10$\% ($\pi$), $\pm 20$\% ($\rho$), $\pm 30$\% ($\omega$), $\pm 50$\% ($\eta$) and  $\pm 100$\% ($\eta^{\prime}$)~\cite{HFAG}. The relative contribution of the unmeasured components of the hybrid model MC are varied within the limits of the full inclusive branching fraction.The inclusive part of the mix uses a model based the SF parameterization in Ref.~\cite{dfn}. The hybrid MC is corrected to match the moments of the $q^2$ and $M_X$ distributions as predicted by the model in Ref.~\cite{Gambino:2007rp}.   The uncertainty in the inclusive component is determined by taking into account the error on the SF parameters and the theoretical and intrinsic uncertainties in the models in Refs.~\cite{dfn, Gambino:2007rp}. We estimate the uncertainty due to the simulation of kaon production in $B \to X_u \ell \nu$ decays ({\it i.e.} gluon splitting into an $s \bar s$ pair), by varying the contribution of events with a kaon by 25\%.  

Systematic errors in the subtraction of the non-$B \overline{B}$ background are dominated by the uncertainty in the relative normalization of the on- and off-resonance data, which is estimated to be a $1$\% error on the continuum yield.  The uncertainty due to mis-tagging is estimated by varying the lower bound on the $M_{\rm{bc}}$ signal region, corresponding to a $10$\% variation in the ratio of good tags to incorrect tags in the signal region~\cite{Urquijo:2006wd}.

The systematic uncertainty due to the overall fit to data for the background contribution normalization is estimated by varying the number of bins used in the fit. The uncertainty due to secondary, cascade $B \to D \to e$ decays is assessed by varying the branching fractions of semileptonic $D$ decays, and $B \to D$ {\it anything} by $\pm1\sigma$~\cite{PDG}. The uncertainty associated to the magnitude of the hadron fake contribution is determined from measurements of $K^0_{\rm S} \to \pi^+ \pi^-$ decays.

To model backgrounds from $B \to D \ell \nu$ and $B \to D^* \ell \nu$ decays we use parameterizations of the form factors based on heavy quark effective theory~\cite{HQET,CLN,GL}.    The $B \to D \ell \nu$ and $B \to D^* \ell \nu$ decay slope parameters  are set to the world averages \cite{HFAG}. The $B \to D^* \ell \nu$ decay parameters, $R_1$ and $R_2$ are set to the most recently measured values\cite{HFAG}. The branching fractions of the $D$ and $D^*$ components are based on Ref.~\cite{PDG}.    For higher mass $D^{**}$ resonances we use the model in Ref.~\cite{LLSW} with the method described in Ref.~\cite{Urquijo:2006wd}.   We adopt the prescription of Ref.~\cite{ref:9} for the non-resonant $B \to D^{(*)} \pi \ell \nu$ decay shapes. The normalization of the narrow resonant $D^{**}$ and non-resonant $D^*\pi$ components are based on values in Ref.~\cite{HFAG}. The remaining unmeasured contribution is matched to the full inclusive rate. To estimate the sensitivity to the rates of the exclusive $B \to X_c \ell \nu$ modes, we adjust their individual branching fractions about their measured uncertainties.  To test the sensitivity to the shape of these contributions, we have varied the form factors for $D^* \ell \nu$, and $D \ell \nu$ decays about their measured uncertainties, and changed the model input parameters that describe the differential decay rates of the resonant $D^{**} \ell \nu$ decays. For the resonant $D^{**} \ell \nu$ decays, we take into account limits from measurements to resonant and non-resonant $D^{(*)} \pi \ell \nu$ states, and full inclusive rates~\cite{matsumoto,HFAG,PDG}. The  systematic uncertainty on the non-resonant $D^{(*)} \pi \ell \nu$ decay modes is estimated as half of the shift between the bounds on the branching fractions. The simulation of QED corrections incurs a negligible systematic error ~\cite{Urquijo:2006wd}.

We estimate the effect of the specific choice of parameters used in training the BDT by varying the pruning technique, the number of trees, and the minimum number of events in each node by 20\% for each respective quantity.

The partial branching fraction for $p^{*B}_{\ell}>1.0$~GeV/$c$ is $ \Delta \mathcal{B} (B \to X_u \ell \nu; p^{*B}_{\ell}>1.0~{\rm GeV}/c) =1.963 \times (1 \pm  0.088_{\rm stat.}  \pm  0.081_{\rm sys.}   )\times 10^{-3}$. A breakdown of the uncertainties is provided in  Table~\ref{tab:br_charmless_sys}.  
\begin{table}[htb]
\caption{Uncertainties in the partial charmless semileptonic branching fraction (in percent).}
\label{tab:br_charmless_sys}
\begin{center}
\begin{tabularx}{0.4\linewidth}{l||c}
\hline \hline
$p^{*B}_{\ell}>1.0$~GeV                              &$\Delta\mathcal{B}/\mathcal{B}$ (\%)  \\ 
\hline\hline
$\mathcal{B}(D^{(*)} \ell \nu)$                                                        & $1.2$                     \\
$(D^{(*)} \ell \nu)$ form factors                                                       & $1.2$                    \\
$\mathcal{B}(D^{**}e\nu)$ \& form factors                      & $0.2$                   \\
\hline
$B \to X_u \ell \nu$ (SF)                                                                & $3.6$                  \\
$B \to X_u \ell \nu$ ($g\to s \bar s$)					    &$1.5$\\
\hline
$\mathcal{B}$($B \to \pi / \rho / \omega \ell \nu$)                                  & $2.3$           \\
$\mathcal{B}$($B \to  \eta,~\eta^{\prime} \ell \nu$)      & $3.2$           \\
$\mathcal{B}$($B \to X_u \ell \nu$) un-meas.                                  & $2.9$           \\
\hline
Cont./Comb.                                                                                   & $1.8$                 \\
Sec./Fakes/Fit.                                                                                & $1.0$                     \\
\hline                                                
PID/Reconstruction                                                                                        & $3.1$               \\
\hline
BDT                                                                 & $3.1$                 \\
\hline
\hline
Systematics                                                                                   & $8.1$                \\
Statistics                                                                                         & $8.8$               \\
\hline
\end{tabularx}
\end{center}
\end{table}
We obtain $|V_{ub}|$ directly from the partial branching fraction using  $|V_{ub}|^2=\Delta \mathcal{B}_{u \ell \nu}/(\tau_{B}\Delta{\cal R})$, where $\Delta {\cal R}$ is the predicted $B \to X_u \ell \nu$ partial rate in the given phase space region, and $\tau_B$ is the average $B$ lifetime~\cite{HFAG}.   Table \ref{tab:vub} lists $|V_{ub}|$ values extracted with the most recent QCD calculations ~\cite{BLNP, DGE, Gambino:2007rp}, where the errors are statistical, systematic, from the error on $m_b$, and theoretical, respectively.  Within their stated theoretical uncertainties, the results in Table \ref{tab:vub} are consistent. 

\begin{table}[htb]
\caption{Values for $|V_{ub}|$ with relative errors (in \%).}
\label{tab:vub}
\begin{center}
\begin{tabularx}{0.5\linewidth}{Xccccc}
\hline \hline
Theory &$|V_{ub}|\times10^{3}$ & Stat. & Sys. & $m_b$ & Th.\\
\hline\hline
 BLNP \cite{BLNP} &$4.37$& $4.3$ & $4.0$ & ${}^{+3.1}_{-2.7}$ &${}^{+4.3}_{-4.0}$\\
\hline
DGE \cite{DGE} &$4.46$& $4.3$ & $4.0$ & ${}^{+3.2}_{-3.3}$ &${}^{+1.0}_{-1.5}$\\
\hline
GGOU \cite{Gambino:2007rp} &$4.41$& $4.3$ & $4.0$ & $1.9$                           &${}^{+2.1}_{-4.5}$\\ 
\hline
\end{tabularx}
\end{center}
\end{table}

%

In summary, a new experimental technique has been employed to measure the branching fraction of inclusive charmless semileptonic $B$ decays ($B \to X_u \ell \nu$) over nearly the full kinematical phase space, resulting in a large reduction on the uncertainty of $|V_{ub}|$.  We provide a more reliable, comprehensive treatment of the many contributions to the signal and background processes, while reducing the experimental (non-model) systematic errors on $|V_{ub}|$ by $\sim30$\%, with respect to Ref.~\cite{Bizjak:2005hn}.   The theoretical uncertainties on $|V_{ub}|$, dominated by the uncertainties on the $b$-quark mass and the shape function, are $40$\% lower in the schemes in Ref. \cite{DGE} and \cite{Gambino:2007rp} and $20$\% lower in the scheme in Ref. \cite{BLNP} than in our previous measurement~\cite{Bizjak:2005hn}.  The SF errors have been almost completely removed from the theoretical extrapolation. The improvement in the uncertainty is primarily due to the increase in the measured phase space, which decreases the power dependence of $|V_{ub}|$ on the $b$-quark mass.  These values  have an overall uncertainty of $\sim 7\%$, competitive with that of the world average~\cite{HFAG}, and in agreement at the $\sim 1 \sigma$ level.  This result increases our confidence in the inclusive determination of $|V_{ub}|$, further highlighting the gap between the inclusive and the exclusive determinations, and with $\sin{2 \phi_1}$.  This is the last measurement of inclusive $|V_{ub}|$ by Belle, using the full $\Upsilon(4S)$ data sample.

We thank the KEKB group for excellent operation of the accelerator, the KEK cryogenics group for efficient solenoid operations, and the KEK computer group and the NII for valuable computing and SINET3 network support.   We acknowledge support from MEXT, JSPS and Nagoya's TLPRC (Japan); ARC and DIISR (Australia); NSFC (China);  DST (India); MOEHRD and KOSEF (Korea); MNiSW (Poland);  MES and RFAAE (Russia); ARRS (Slovenia); SNSF (Switzerland);  NSC and MOE (Taiwan); and DOE (USA).


\begin{thebibliography}{99}
\bibitem{notation}
This angle is also known as $\beta$ in the literature.
\bibitem{cc} 
Throughout this paper, the inclusion of the charge conjugate mode decay is implied unless otherwise stated.
\end{thebibliography}
\end{document}